\documentclass[conference]{IEEEtran}
\IEEEoverridecommandlockouts
\usepackage{cite}
\usepackage{amsmath,amssymb,amsfonts}
\usepackage{algorithmic}
\usepackage{graphicx}
\usepackage{subfig}   % IEEE-compatible
\usepackage{textcomp}
\usepackage{xcolor}
\usepackage{hyperref}

\def\BibTeX{{\rm B\kern-.05em{\sc i\kern-.025em b}\kern-.08em
    T\kern-.1667em\lower.7ex\hbox{E}\kern-.125emX}}
\begin{document}

\title{Joint Simplicial Complex Learning \\ via Binary Linear Programming
%{\footnotesize \textsuperscript{}}
\thanks{This publication is part of the project CYCLONE with file number OCENW.M.24.255 of the research programme NWO Open Competition Domain Science – M, which is (partly) financed by the Dutch Research Council (NWO) under the grant with ID https://doi.org/10.61686/UBZOA14639}
}

\author{Varun Sarathchandran and Geert Leus\\
%\textit{Department of Microelectronics} \\
\textit{Delft University of Technology}, Delft, The Netherlands \\
\texttt{ \{V.Sarathchandran;G.J.T.Leus\}@tudelft.nl } }

\maketitle

\begin{abstract}
Learning the topology of higher-order networks from data is a fundamental challenge in many signal processing and machine learning applications. Simplicial complexes provide a principled framework for modeling multi-way interactions, yet learning their structure is challenging due to the strong coupling across different simplicial levels imposed by the inclusion property. In this work, we propose a joint framework for simplicial complex learning that enforces the inclusion property through a linear constraint, enabling the formulation of the problem as a binary linear program. The objective function consists of a combination of smoothness measures across all considered simplicial levels, allowing for the incorporation of arbitrary smoothness criteria. This formulation enables the simultaneous estimation of edges and higher-order simplices within a single optimization problem. Experiments on simulated and real-world data demonstrate that the proposed joint approach outperforms hierarchical and greedy baselines, while more faithfully enforcing higher-order structural priors.
\end{abstract}


\section{Introduction}
Higher-order networks have gained increasing attention as they capture interactions beyond pairwise relationships, which are prevalent in many real-world systems. In domains such as biology, epidemiology, and the social sciences, interactions often occur among groups of entities rather than simple pairs, motivating representations that go beyond standard graphs~\cite{bick2023higher}. One such higher-order network model is the simplicial complex (SC), which explicitly represents multi-way interactions while enforcing an inclusion structure across different interaction orders~\cite{BarbarossaTopologicalComplexes}.

For effective signal processing on a network, the underlying topology must be known. In many applications, however, it must be inferred from data. While graph topology identification is well studied~\cite{Mateos2018ConnectingProcessing}, learning SCs from data remains relatively unexplored. A key challenge in SC learning is that connectivity across different simplicial levels is strongly coupled through the inclusion property, which makes direct, one-shot learning difficult.

One way recent work addresses this challenge is by adopting hierarchical approaches, in which lower-order simplices are inferred first, followed by higher-order simplices. For example, in~\cite{BarbarossaTopologicalComplexes}, triangles are inferred from edge signals given a known edge structure by minimizing the curl of the edge flow. In~\cite{GurugubelliSimplicialSampling}, simplicial signals are modeled as random variables, leading to a formulation that focuses on learning triangles from edge signals under a smoothness prior. A related probabilistic framework is proposed in~\cite{Sardellitti2023ProbabilisticComplexes}. While effective in certain settings, these hierarchical approaches decouple the estimation of different simplicial levels and do not fully exploit the structural dependencies across levels.

Non-hierarchical approaches are however more preferred because they allow the connectivity on higher orders to also guide the lower-order connectivity, improving performance. An interesting example is~\cite{BuciuleaLEARNINGDATA}, where a Volterra-based model is used~\cite{Self-DrivenPrediction}. This method uses group sparsity penalties to couple the first and second-order topology by linking the cost of activating a triangle to that of activating its constituent edges. However, the inclusion property is enforced only implicitly through penalization, which does not guarantee inclusion. More recently, a greedy approach is proposed in~\cite{BuciuleaLearningApproach}, where a bilinear constraint is introduced to enforce inclusion. This formulation requires alternating optimization between edges and triangles, and the constraint is incorporated as a penalty term in the objective. As we show experimentally, when this penalty is sufficiently large, the method effectively reduces to a hierarchical procedure.

In contrast, we propose a joint, one-shot framework for learning SCs by explicitly enforcing the inclusion property through a linear constraint that couples consecutive simplicial levels. This formulation allows SC learning to be posed as a binary linear program, enabling simultaneous estimation of edges and higher-order simplices within a single optimization problem. Through simulations and experiments on real data, we demonstrate the superior performance of the proposed joint approach compared to hierarchical and greedy baselines.

\section{Preliminaries}

We begin with a conceptual description of simplicial complexes and signals \cite{SchaubSignalBeyond}\cite{ISUFI2025109930}, where we will also introduce notation. We then present several simplicial Laplacians, their algebraic representation and corresponding smoothness measures, which relate simplicial topology and signals. 
\subsection{Simplicial Complexes and Signals} 
Given a set of nodes $\mathcal{N}$, a $k$-simplex is defined as a subset of $\mathcal{N}$ with cardinality $k+1$. Nodes are thus 0-simplices, edges are 1-simplices, and triangles are 2-simplices. A simplicial complex is a collection of $k$-simplices, subject to an inclusion property. For every $k$-simplex in the complex, its $(k-1)$-simplices must also be included in the complex. Thus, for instance, all edges of a triangle must also be included in the complex. The $(k-1)$-simplices of a $k$-simplex are called its faces. Similarly, a $(k+1)$-simplex that includes a $k$-simplex is called its coface. Two $k$-simplices are lower-adjacent if they share a common face, and are upper-adjacent if they share a common coface. The upper and lower adjacencies induce the upper and lower neighborhoods of a $k$-simplex. A $k$-simplicial signal, formed by collecting a scalar value over each $k$-simplex, is defined as a vector $\mathbf{x}_k \in \mathbb{R}^{N_k}$, where $N_k$ is the number of $k$-simplices in the complex. Extending to a vector per simplex, we get $\mathbf{X}_k \in \mathbb{R}^{N_k \times F_k}$, where $F_k$ is the number of features on the $k$th level. In this work, $[\mathbf{x}_k]_i$ denotes the $i$th entry of the vector $\mathbf{x}_k$, and $[\mathbf{X}_k]_{i,j}$ denotes the $(i,j)$th entry of the matrix $\mathbf{X}_k$.

%\subsection{The Hodge Laplacian and its Decomposition}

\subsection{Simplicial Laplacians and Smoothness} \label{sec:smoothness}

%To enable a mathematical representation of simplicial complexes and signals, we now introduce the Hodge Laplacian and its associated decomposition. 
%The mapping between $(k-1)$-simplices and $k$-simplices is given by the incidence matrix $\mathbf{B}_k \in \mathbb{R}^{N_{k-1} \times N_{k}}$. The $i^{\text{th}}$ column of $\mathbf{B}_k$ reveals the faces of the $i^{\text{th}}$ $k$-simplex through $k+1$ nonzero entries taking values in $\{+1,-1 \}$, with the sign indicating an appropriate orientation. 
%To enable a mathematical representation of simplicial complexes and relate simplicial signals to them, we now introduce several simplicial Laplacians. 
To mathematically represent simplicial complexes and establish a link with simplicial signals, we define several simplicial Laplacians.
The linear map from $(k-1)$-simplices to $k$-simplices is represented by the incidence (boundary) matrix $\mathbf{B}_k \in \mathbb{R}^{N_{k-1} \times N_k}$. Each column of $\mathbf{B}_k$ corresponds to a $k$-simplex, and each row
corresponds to a $(k-1)$-simplex. The entry $[\mathbf{B}_k]_{i,j}$ is nonzero if
and only if the $i$th $(k-1)$-simplex $\sigma_{k-1}^{(i)}$ is a face of the $j$th $k$-simplex
$\sigma_{k}^{(j)}$. In this case, $[\mathbf{B}_k]_{i,j} \in \{+1,-1\}$, where the sign is
determined by whether the orientation induced on $\sigma_{k-1}^{(i)}$ from $\sigma_{k}^{(j)}$ agrees or disagrees with the chosen reference orientation of $\sigma_{k-1}^{(i)}$. Each column of $\mathbf{B}_k$ contains exactly $k+1$ nonzero entries. A simplicial complex of order $K$ (the highest level in the complex) can be defined via the Hodge Laplacians \cite{Lim2020HodgeGraphs}:
\begin{equation}
    \mathbf{L}_k = \mathbf{B}_k^\top   \mathbf{B}_k + \mathbf{B}_{k+1}   \mathbf{B}_{k+1}^\top, \quad k=1,\dots, K-1.
\end{equation}
The first term, the lower Laplacian $\mathbf{L}_k^{\rm low}$, encodes the lower neighbourhood, and the second term, the upper Laplacian $\mathbf{L}_k^{\rm up}$ encodes the upper neighbourhood at level $k$.
For $k=0$ ($k=K$), the Laplacian contains only the second (first) term. The inclusion property induces the constraint $\mathbf{B}_k \mathbf{B}_{k+1} = \mathbf{0}$. 

The Laplacians defined above are commonly used to quantify smoothness measures of simplicial signals, thereby providing a principled way to relate observed signals to the underlying topology. In particular, the smoothness of node signals $\mathbf{X}_0$ over the edges can be measured using the graph Laplacian $\mathbf{L}_0$ via the quadratic form $\operatorname{tr}(\mathbf{X}_0^\top \mathbf{L}_0 \mathbf{X}_0)$.
This expression captures variations of the signal across adjacent nodes and serves as a widely used prior for learning the graph structure from data \cite{Mateos2018ConnectingProcessing}\cite{Kalofolias2016HowSignals}\cite{DongLearningRepresentations}. 

For triangles, the Laplacian $\mathbf{L}_1^{\rm up}$ has been widely used to define a notion of the curl of edge signals ${\bf X}_1$ via the quadratic form $\operatorname{tr}(\mathbf{X}_1^\top \mathbf{L}_1^{\rm up} \mathbf{X}_1)$. This measure captures the conservation of edge signals around a triangle via an orientation-consistent signed sum of its constituent edge signals \cite{BarbarossaTopologicalComplexes}\cite{BuciuleaLearningApproach}. 

This curl-based formulation, however, represents only one notion of edge-signal smoothness. While it captures conservation of the edge signal around a triangle, it does not directly quantify how similar the individual edge values are within that triangle. This observation motivates an alternative notion of smoothness that explicitly measures similarity
%\footnote{More precisely, this quantity measures dissimilarity (or variation). We refer to it as a similarity measure to distinguish it from the node-level smoothness measure.} 
among the edges of a triangle, or in general, the faces of a $k$-simplex, which we introduce next.

Let $\mathcal{F}(\sigma^{(i)}_k)$ denote the set of faces of the $i$th $k$-simplex $\sigma^{(i)}_k$.
We then define the similarity for $k$-simplices as
\begin{equation}
\textstyle     \sum_{i=1}^{N_k}
\ \sum_{\substack{f,g \in \mathcal{F}(\sigma^{(i)}_k)\\ f<g}}
\| [\mathbf{X}_{k-1}]_{f,:} - [\mathbf{X}_{k-1}]_{g,:}\|^2_2.
\end{equation}
%{\color{red} GL: I think there might be an issue here actually with assuming $f<g$. For instance, for edges every pair of nodes is considered twice I think. At least if you use the Laplacian interpretation. Am I correct?}
When $k=1$, this expression reduces to the standard graph Laplacian quadratic form for node signals, as it sums squared differences between the two nodes of each edge. Consequently, the proposed measure generalizes classical node-signal smoothness on graphs to higher-order simplices. This similarity measure also admits a Laplacian interpretation. Let $\mathbf{L}^\Delta_{k}$ be the Laplacian of the complete graph on the $k+1$ faces of a $k$-simplex, and $\mathbf{C}_k^{(i)} \in \{0,1\}^{(k+1) \times N_{k-1}}$ be a selection matrix which selects the faces of $\sigma^{(i)}_k$. Then, the similarity measure can be equivalently written as 
\begin{equation}    
\textstyle
\operatorname{tr}(\mathbf{X}_{k-1}^\top ( \sum_{i=1}^{N_k} {\mathbf{C}^{(i)}_k}^\top \mathbf{L}^\Delta_{k} \mathbf{C}_k^{(i)} ) \mathbf{X}_{k-1} ),
\label{eq:similarity_laplacian}
\end{equation}
which naturally leads to a Laplacian interpretation with $\mathbf{L}^{\text{sim}}_{k} = \sum_{i=1}^{N_k}  {\mathbf{C}^{(i)}_k}^\top \mathbf{L}^\Delta_{k} \mathbf{C}^{(i)}_k$. In the following sections, we show how priors based on these smoothness notions can be leveraged to learn the underlying SC topology. 

\section{Problem Formulation}

In this section, we formally state the problem of SC recovery. While the proposed framework applies to SCs of any order, we restrict the discussion to edges and triangles for clarity. Given a set of nodes $\mathcal{N}$, nodal observations $\mathbf{X}_0 \in \mathbb{R}^{N_0 \times F_0}$, and edge-level signals $\mathbf{\bar{X}}_1 \in \mathbb{R}^{\bar{N}_1 \times F_1}$ defined over the full candidate edge set of size $\bar{N}_1 = \binom{N_0}{2}$, our objective is to recover the underlying SC topology. 
%Here $\bar{N}_1 = \binom{N_0}{2}$ denotes the size of the candidate edge set, that is, the set of all possible edges. 
In this work, we focus on unweighted SCs, and leave weighted SCs for future work. Specifically, we seek to identify the incidence structure of edges and second-order simplices (triangles), encoded through the incidence matrices $\mathbf{B}_1$ and $\mathbf{B}_2$, subject to the inclusion property. To this end, we impose prior assumptions that relate the observed simplicial signals to the underlying topology, and exploit these relationships to infer the unknown incidence structure. This leads to the following general optimization framework:
\begin{subequations}\label{eq:general_SC_problem}
\begin{align} %\label{eq:general_SC_problem}
\min_{\mathbf{B}_1,\mathbf{B}_2}\quad 
& f_1(\mathbf{B}_1,\mathbf{X}_0) + f_2(\mathbf{B}_2,\mathbf{\bar{X}}_1) 
\label{eq:general_SC_problem_obj} 
\\
\text{s.t.}\quad 
& \mathbf{B}_1 \in \mathcal{B}_1, \ \ \mathbf{B}_2 \in \mathcal{B}_2,
, \ \ 
 \mathbf{B}_1 \mathbf{B}_2 = \mathbf{0}. 
\label{eq:general_SC_problem_constraints}
\end{align}
\end{subequations}
Here, the functions $f_1$ and $f_2$ in~\eqref{eq:general_SC_problem_obj} encode priors that couple the observed simplicial signals with the underlying topology. The sets $\mathcal{B}_1$ and $\mathcal{B}_2$ in~\eqref{eq:general_SC_problem_constraints} denote the feasible incidence matrices for edges and triangles, respectively, which also includes the criterion that a minimum number of edges and triangles exist. The first two constraints in~\eqref{eq:general_SC_problem_constraints} restrict the incidence matrices to these feasible sets while the last constraint in~\eqref{eq:general_SC_problem_constraints} enforces the simplicial inclusion property.

Note that the availability of edge-level signals in~\eqref{eq:general_SC_problem} does not imply prior knowledge of the edge structure. Instead, signals are defined over all possible node pairs $\bar{N}_1$. Our formulation is agnostic to how edge-level signals are obtained, as long as they are available across the entire candidate edge set. In practice, edge-level signals may arise in different ways. In this work, we consider two representative scenarios. In one, edge signals are not directly observed and are instead constructed from node-level measurements using a permutation-invariant mapping, such as in \cite{Self-DrivenPrediction}. In the other, edge signals are directly observed and available over the entire candidate edge set. Specific details will follow in our experiment section. 

\section{Binary Linear Program Formulation}

The formulation in~\eqref{eq:general_SC_problem} is intentionally general. In this section, we describe a concrete method that enables joint recovery of the SC topology via a binary linear program. This formulation raises three key challenges: (i) enforcing feasibility of the incidence matrices $\mathbf{B}_1$ and $\mathbf{B}_2$, (ii) ensuring the inclusion property is satisfied, and (iii) designing suitable objective functions $f_1$ and $f_2$ that couple topology and observed signals. We address these challenges in the subsequent paragraphs.

Following the simplex selection perspective introduced in \cite{BuciuleaLearningApproach}, we avoid optimizing directly over the incidence matrices $\mathbf{B}_1$ and $\mathbf{B}_2$. Instead, we operate in the space of fully connected SCs, characterized by the full incidence matrices $\bar{\mathbf{B}}_1 \in \mathbb{R}^{N_0 \times \bar{N}_1}$ and $\bar{\mathbf{B}}_2 \in \mathbb{R}^{\bar{N}_1 \times \bar{N}_2}$ that enumerate all possible edges and triangles, with $\bar{N}_1 = \binom{N_0}{2}$ and $\bar{N}_2 = \binom{N_0}{3}$.
From the full SC, binary selection vectors $\mathbf{s}_1 \in \{0,1\}^{\bar{N}_1}$ and $\mathbf{s}_2  \in \{0,1\}^{\bar{N}_2}$ are then used to pick active edges and triangles. We then work entirely in the space of $\mathbf{s}_1$ and $\mathbf{s}_2$, which avoids having to directly optimize over $\mathbf{B}_1$ and $\mathbf{B}_2$. Ensuring that a minimum number of edges and triangles are selected boils down to constraining  $\mathbf{s}_1$ and $\mathbf{s}_2$ to have a minimum number of non-zero entries. 

Designing an effective method to enforce the inclusion property is key to allow joint recovery of the SC topology. We achieve this by introducing a linear constraint that couples the edge and triangle selection variables, given by
\begin{equation}
    \mathbf{s}_1 \geq \alpha \bar{\mathbf{B}}_2^{+} \mathbf{s}_2.
    \label{eq:linearSCconstraint}
\end{equation}
Here, $\bar{\mathbf{B}}_2^{+}$ denotes the unoriented edge-to-triangle incidence matrix. The vector $\bar{\mathbf{B}}_2^{+}\mathbf{s}_2$ counts, for each candidate edge, the number of selected triangles that it is included in. For sufficiently small $\alpha > 0$, this constraint ensures that whenever an edge participates in at least one selected triangle, the corresponding entry of $\mathbf{s}_1$ must be active, thereby enforcing simplicial inclusion. This linear constraint enables joint recovery of edges and triangles within a unified framework. This contrasts with other coupling mechanisms that lead to nonconvex formulations and require alternating schemes. 
%Extending \eqref{eq:linearSCconstraint} to weighted SCs, where $\mathbf{s}_1,\mathbf{s}_2 \in [0,1]$, requires careful treatment. In the weighted setting, directly applying \eqref{eq:linearSCconstraint} imposes a lower bound on each edge weight which is $\alpha$ times the total weight of its incident triangles. The implications of such a coupling between simplex weights are not immediately clear and warrant separate study, which we leave for future work.

What remains is to specify the functions $f_1$ and $f_2$ which encode prior assumptions relating signals to the underlying topology. Since we recast our optimization problem with $\mathbf{s}_1$ and $\mathbf{s}_2$, the objective in \eqref{eq:general_SC_problem} is redefined as $f_1(\mathbf{s}_1,\mathbf{X}_0) + f_2(\mathbf{s}_2,\mathbf{\bar{X}}_1)$. In this work, we restrict attention to functions that are linear in $\mathbf{s}_1$ and $\mathbf{s}_2$, and thus we can write $f_1(\mathbf{s}_1,\mathbf{X}_0)= \mathbf{h}_1^\top \mathbf{s}_1$  and $f_2(\mathbf{s}_2,\mathbf{\bar{X}}_1) = \mathbf{h}_2^\top \mathbf{s}_2$. Here, each entry of $\mathbf{h}_1$ ($\mathbf{h}_2$) assigns a cost to a candidate edge (triangle) based on the associated node (edge) signals. This choice, along with the linear simplicial inclusion constraint, enables posing SC topology learning as a binary linear program, while remaining sufficiently expressive to capture a wide range of signal-topology relationships, including those specified in Section~\ref{sec:smoothness}.

To make this connection explicit, we now show how smoothness priors can be expressed within our selection-variable formulation. Since we optimize over variables defined on the full candidate complex, all Laplacian-based smoothness measures are written in this full space, with the selection vectors activating only the simplices that are present. In particular, the node signal smoothness $\operatorname{tr}(\mathbf{X}_0^\top \mathbf{\bar{L}}_0 \mathbf{X}_0)$, with $\mathbf{\bar{L}}_0 = \bar{\mathbf{B}}_1 \operatorname{diag}(\mathbf{s}_1) \bar{\mathbf{B}}_1^\top$, admits the linear form ${\operatorname{diag}(\bar{\mathbf{B}}_1^\top \mathbf{X}_0\mathbf{X}_0^\top \bar{\mathbf{B}}_1)}^\top \mathbf{s}_1$. Accordingly, we define $\mathbf{h}_1 = {\operatorname{diag}(\bar{\mathbf{B}}_1^\top \mathbf{X}_0\mathbf{X}_0^\top \bar{\mathbf{B}}_1)}$ for edge recovery throughout this manuscript.
For triangles, we consider two alternatives: (i) as a first option, we consider the curl-based smoothness, which admits a linear representation of the form $\operatorname{tr}(\mathbf{\bar{X}}_1^\top \bar{\mathbf{L}}_1^{\rm up} \mathbf{\bar{X}}_1)
= \operatorname{diag}\!\big(\bar{\mathbf{B}}_2^\top \mathbf{\bar{X}}_1\mathbf{\bar{X}}_1^\top \bar{\mathbf{B}}_2\big)^\top \mathbf{s}_2$, with $\bar{\mathbf{L}}_1^{\rm up} = \bar{\mathbf{B}}_2 \operatorname{diag}(\mathbf{s}_2)\bar{\mathbf{B}}_2^\top$. This yields $\mathbf{h}_2 = \operatorname{diag}\!\big(\bar{\mathbf{B}}_2^\top \mathbf{\bar{X}}_1\mathbf{\bar{X}}_1^\top \bar{\mathbf{B}}_2\big)$. (ii) As a second choice, we consider the proposed similarity-based smoothness measure in~\eqref{eq:similarity_laplacian}. This is more easily expressed at the level of individual simplices and leads to  $[\mathbf{h}_2]_t = \operatorname{tr} (\bar{\mathbf{X}}_1^\top \bar{\mathbf{C}}_2^{(t)\top} \mathbf{L}^\Delta_2 \bar{\mathbf{C}}_2^{(t)} \bar{\mathbf{X}}_1 )$, where $\bar{\mathbf{C}}_2^{(t)} \in {\mathbb R}^{3 \times \bar{N}_1}$ selects the three edges related to triangle $t$ from the full candidate set of edges. Both these smoothness measures will be considered to recover triangles in all our experiments.

We are now ready to present our framework. We posit that the SC topology can be jointly recovered from simplicial signals via the following binary linear program:
  \begin{subequations}\label{eq:BLP_SC_learning}
\begin{align}
\min_{\mathbf{s}_1,\mathbf{s}_2}\quad 
& \mathbf{h}_1^\top \mathbf{s}_1 + \mathbf{h}_2^\top \mathbf{s}_2
\label{eq:BLP_SC_learning_obj} \\[1mm]
\text{s.t.}\quad 
& \mathbf{s}_1 \geq \alpha \bar{\mathbf{B}}_2^{+}\mathbf{s}_2
\label{eq:BLP_SC_learning_inclusion} \\
& \mathbf{s}_1 \in \{0,1\}^{\bar N_1}, \quad 
  \mathbf{s}_2 \in \{0,1\}^{\bar N_2}
\label{eq:BLP_SC_learning_binary} \\
& \mathbf{1}^\top\mathbf{s}_1 \geq C_1, \quad 
  \mathbf{1}^\top\mathbf{s}_2 \geq C_2.
\label{eq:BLP_SC_learning_cardinality}
\end{align}
\end{subequations}
Constraint~\eqref{eq:BLP_SC_learning_inclusion} enforces the simplicial inclusion property. Constraint~\eqref{eq:BLP_SC_learning_binary} restricts the formulation to unweighted simplicial complexes. Finally, we impose lower bounds on the number of selected edges and triangles through the cardinality constraint~\eqref{eq:BLP_SC_learning_cardinality}. Since the selection variables are binary, this linear constraint is also equivalent to an $\ell_0$-norm constraint. 

%Numerical results follow in the experiments section. 

\section{Experiments}
\begin{figure*}[t]
    \centering
    \includegraphics[width=\linewidth]{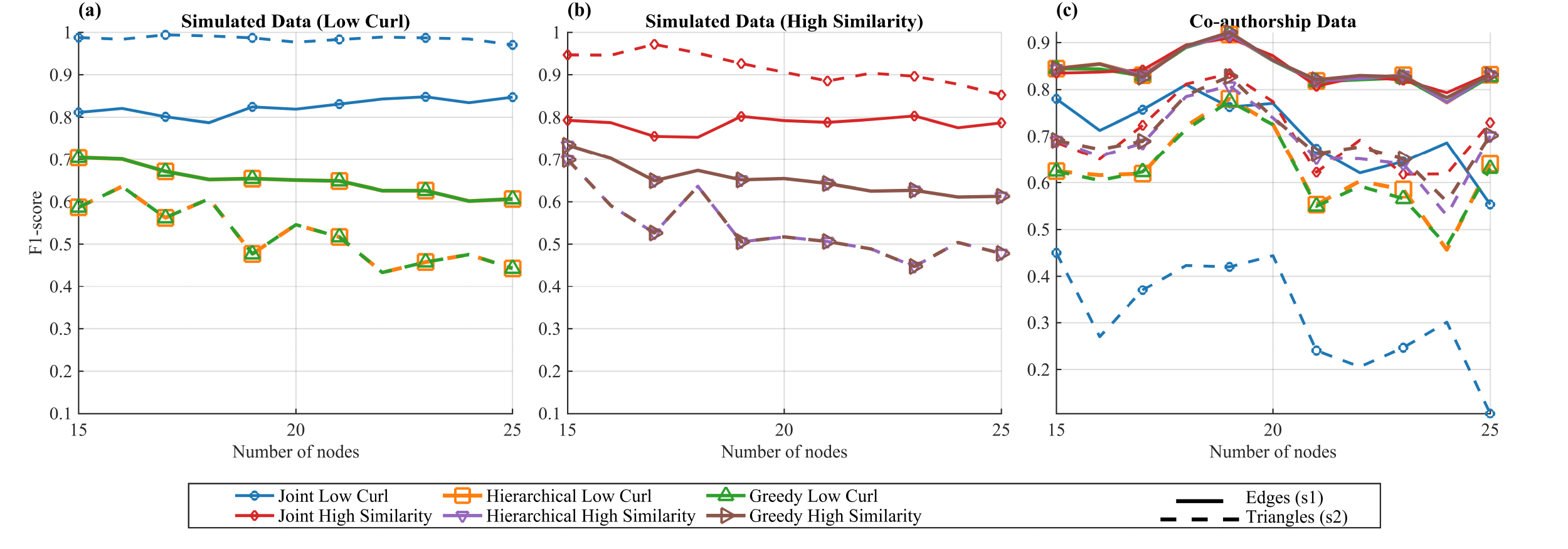}
     \caption{The f1 scores of methods on edge and triangle detection on simulated and real data. All methods rely on node signal smoothness to identify edges, and thus labels specify the type of smoothness used to identify triangles. (a) Performance on simulated data with edge signals having low curl on triangles. (b) Performance on simulated data with similar edge signals on a triangle. (c) Performance of all methods with both smoothness types on real co-authorship data. Results are aggregated over 10 random realizations.}
    \label{fig:results}
\end{figure*}
In this section, we describe our baselines, and present results on simulated and real data.
\subsection{Baselines}
Our first baseline is a hierarchical approach that estimates the edge and triangle structures sequentially. Specifically, the edge selection vector $\mathbf{s}_1$ is first estimated, after which triangles are inferred by restricting attention to those that are feasible given the recovered edges. An estimate of the edge selection vector $\hat{\mathbf{s}}_1$ is obtained by solving:
\begin{subequations}
    \begin{align}
\min_{\mathbf{s}_1}\quad  & \mathbf{h}_1^\top \mathbf{s}_1 \\
\text{s.t.}\quad 
& \mathbf{1}^\top\mathbf{s}_1 \geq C_1, \quad 
\mathbf{s}_1 \in \{0,1\}^{\bar N_1}.
\end{align}
\end{subequations}
Next, we restrict $\mathbf{s}_2$ to the feasible triangle set $\mathcal{T}(\hat{\mathbf{s}}_1)$, consisting of all triangles whose three edges belong to $\hat{\bf s}_1$,
%:= \{\, t \in \{1,\dots,\bar N_2\}: \text{all three edges of triangle $t$ are selected in $\hat{\mathbf{s}}_1$}\},
and solve,
\begin{subequations}\label{eq:baseline_s2_restricted}
\begin{align}
\min_{\mathbf{s}_2}\quad & \mathbf{h}_2^\top \mathbf{s}_2
\label{eq:baseline_s2_restricted_obj}\\
\text{s.t.}\quad 
& [{\mathbf{s}_2}]_t = 0,\quad \forall t \notin \mathcal{T}(\hat{\mathbf{s}}_1)
\label{eq:baseline_s2_restricted_support}\\
&\mathbf{1}^\top \mathbf{s}_2 \ge C_2,\quad 
\mathbf{s}_2 \in \{0,1\}^{\bar N_2}.
\label{eq:baseline_s2_restricted_card}
\end{align}
\end{subequations}
It is important to note that if $C_2$ triangles are not feasible, then the hierarchical method is forced to relax the constraint and choose as many triangles as feasible. Under node-signal smoothness for edge recovery and low-curl edge signals for triangle recovery, the hierarchical baseline reduces to the combination of~\cite{Chepuri2016LEARNINGPRIOR} followed by~\cite{BarbarossaTopologicalComplexes}. 

Our second baseline is based on the greedy approach introduced in \cite{BuciuleaLearningApproach}.  In that work, a bilinear simplicial inclusion constraint is presented,  of the form, $( \mathbf{1} - \mathbf{s}_1 )^\top \bar{\mathbf{B}}_2^{+} \mathbf{s}_2 = 0$, which is then relaxed to a penalty cost in the objective. The problem is solved alternately between $\mathbf{s}_1$ and $\mathbf{s}_2$. Specifically, the method considers the following optimization problem:
\begin{subequations}
\begin{equation}
\begin{split}
\min_{\mathbf{s}_1,\mathbf{s}_2}\quad 
&  \|\mathbf{s}_1\|_0 +  \|\mathbf{s}_2\|_0
+ \mathbf{h}_1^\top \mathbf{s}_1 
+ \mathbf{h}_2^\top \mathbf{s}_2 \\
& \quad + \gamma (\mathbf{1}-\mathbf{s}_1)^\top 
\bar{\mathbf{B}}_2^{+} \mathbf{s}_2
\end{split}\label{eq:greedy}
\end{equation}
\begin{align}
\text{s.t.}\quad 
& \mathbf{s}_1 \in \{0,1\}^{\bar N_1}, \quad 
  \mathbf{s}_2 \in \{0,1\}^{\bar N_2} \\
& \|\mathbf{s}_1\|_0 \ge C_1, \quad 
  \|\mathbf{s}_2\|_0 \ge C_2.
\end{align}
\end{subequations}
The formulation in~\cite{BuciuleaLearningApproach} also assumes node-signal smoothness and low curl edge signals. A key difference is that~\cite{BuciuleaLearningApproach} assumes a subset of edges are known, and signals are observed on this subset only. We adapt their method to our setting, where edge signals are available over the full candidate edge set and no prior knowledge of the edge structure is assumed.

% In all baselines and experiments, we assume that node signals are smooth over the underlying graph. For triangle selection, we consider two choices for $\mathbf{h}_2$. The first is the low-curl assumption on edge signals. The second is our proposed similarity measure detailed in \eqref{eq:similarity_laplacian}, which measures variations between the three edges of a triangle. 

For all methods, we set $C_1$ and $C_2$ to their ground truth values. All binary linear programs are solved using the branch and bound method present in the MATLAB optimization toolbox~\cite{wolsey1999integer}. For further details, refer to our \href{https://github.com/VarunSarathchandran/Joint-SC-Learning.git}{GitHub repository}.

\subsection{Numerical Results}

%We present results on simulated data, followed by real data. We simulate SCs and signals such that node-level signals are smooth over the underlying graph. For edge signals, we consider two alternative models: (i) edge signals with low curl over triangles, and (ii) edge signals that have high similarity within each triangle.

An overview of the data generation pipeline for our synthetic experiments is as follows. We first fix the number of nodes and sample an Erdős-Rényi graph with a certain edge probability. Smooth node signals are then generated by filtering white noise through the graph Laplacian $\mathbf{L}_0$. From the set of feasible triangles induced by the graph, half are selected at random. Low curl edge signals are then also generated similarly via $\mathbf{L}_1^{\rm up}$. Edge signals which have high similarity scores are generated via $\mathbf{L}^{\text{sim}}_{2}$. For the low-curl based simulations, the set-up is exactly as used in \cite{BuciuleaLearningApproach}. Note here that edge signals are independent from node signals, and are observable across all node-pairs.

Figures~\ref{fig:results}(a) and~\ref{fig:results}(b) report results on the simulated data.
%under the low-curl and edge-smoothness assumptions, respectively. 
Across all graph sizes, the proposed joint method consistently outperforms both the hierarchical and greedy baselines in terms of f1 score of edge and triangle detection. The superior performance of the proposed joint method highlights the benefit of solving for all levels of the SC simultaneously, allowing edge detection to be informed by the selected triangles.

The greedy and hierarchical baselines exhibit identical performance in these experiments. This behavior can be explained by the effect of the inclusion mechanism in the greedy formulation. When $\gamma$, the weight on the inclusion penalty in~\eqref{eq:greedy}, is sufficiently large to prevent violations, the greedy method never selects an infeasible triangle owing to its large cost, causing it to select exactly the same triangles as the hierarchical method. As long as at least $C_2$ feasible triangles are available, both methods select the same set of triangles, leading to identical results. 

Differences between the greedy and hierarchical approaches arise only when fewer than $C_2$ feasible triangles exist. In this case, the hierarchical method relaxes the cardinality constraint and selects all feasible triangles, whereas the greedy method enforces exactly $C_2$ selections, activating some infeasible triangles. This, in turn, can propagate errors to the edge selection in the subsequent iteration. Such situations do not occur in the simulated experiments considered here, explaining the observed equivalence between the two baselines. In contrast, the proposed joint method optimizes edges and triangles simultaneously under the signal-based objective and the inclusion constraint. Rather than prioritizing the prior for edges, and then mostly restricting attention to feasible triangles, the joint formulation attempts to simultaneously select a set of edges and triangles that achieve the lowest combined cost.

Next, we report results on a real-world co-authorship network dataset following~\cite{Tang2022LearningPrior}. Nodes represent authors, and node signals correspond to the frequency of keywords used in their publications. Edges and triangles are formed between authors who have co-authored the same paper. Edge-level signals for all node pairs are constructed using an element-wise minimum of the corresponding node signals. Before discussing the results, we highlight an important property of this dataset. Empirically, we observe that edge signals are not low curl. Instead, in most realizations, edge signals exhibit a higher degree of similarity across triangles.

%We evaluate the three methods under both edge signal assumptions (node signals are assumed to be smooth in all cases). 
From Figure \ref{fig:results}(c), when similarity is assumed, the joint method consistently outperforms the hierarchical and greedy baselines in triangle recovery, while achieving comparable performance on edges. In cases where the joint method underperforms, we find that the edge signals deviate from the similarity assumption. Under the low-curl assumption, the joint method performs worst, which is expected given that the data does not satisfy this prior. When the underlying data respects the prior, the joint method outperforms the baselines. 

Interestingly, the baselines perform well even when the assumed edge signal prior is violated. This exposes a key distinction between the approaches: both the greedy and hierarchical methods primarily rely on feasibility. The hierarchical method exploits the prior only when more than $C_2$ feasible triangles exist, which is rarely the case in our real data. Similarly, the greedy method enforces feasibility first: when fewer than $C_2$ feasible triangles are available, it fills the remaining selections with infeasible triangles that best satisfy the assumed prior, and when more than $C_2$ feasible triangles exist, it effectively reduces to the hierarchical strategy. 
%As a result, both baselines select predominantly feasible triangles regardless of whether they meaningfully satisfy the assumed signal prior. 
In contrast, the joint method consistently enforces the signal-based prior during triangle selection, and is therefore the only approach that reflects whether the assumed edge signal model is compatible with the data.

\section{Conclusion}
In this work, we introduced a framework for learning SCs that enables joint estimation through a binary linear program. The proposed approach captures structural relationships better, leading to superior recovery of both edges and triangles in simulated experiments. We also introduced a novel smoothness measure that captures similarity among edges within a triangle, which can be easily incorporated into the proposed framework. Experiments on real data further show that existing methods primarily rely on the feasibility of higher-order simplices, whereas the joint formulation explicitly enforces signal-based priors during their selection. 

%\printbibliography
\bibliographystyle{IEEEtran}
\bibliography{references-2}

\end{document}